\documentstyle[aps,epsfig,a4]{revtex}
\begin{document}
\pagestyle{myheadings}\markright{$J/\Psi$ suppression in an expanding hadron gas}%
\draft%

\title{$J/\Psi$ suppression in an expanding hadron gas
\thanks{ Talk based on the work done in collaboration with L.Turko \protect\cite{Prorok:2000kv}.}
\thanks{Presented at the XLI Cracow School of Theoretical Physics "Fundamental Interactions",
Zakopane, Poland, June 2-11 2001.}}
\author{Dariusz Prorok}
\address{Institute of Theoretical Physics, University of
Wroc{\l}aw,\\ Pl. Maksa Borna 9, 50-204  Wroc{\l}aw, Poland}
\date{June 5, 2001}
\maketitle
\begin{abstract}
A model for $J/\Psi$ suppression at a high energy heavy ion
collision is presented. The main (and the only) reason for the
suppression is $J/\Psi$ inelastic scattering within hadron
matter. The hadron matter is in the form of a multi-component
non-interacting gas. Also the evolution of the gas, both
longitudinal and transverse, is taken into account. It is shown
that under such circumstances and with $J/\Psi$ disintegration in
nuclear matter added, $J/\Psi$ suppression evaluated agrees well
with NA38 and NA50 data.
\end{abstract}
\section {Introduction }
\label{intro}

On the 10th of February 2000 CERN announced the discovery of the
novel state of matter --- the so-called quark-gluon plasma (QGP).
The existence of the QGP has been predicted upon lattice QCD
calculations (for a review see \cite{Karsch:2001vs;2000vy} and
references therein) and the critical temperature $T_{c}$ for the
ordinary hadron matter-QGP phase transition has been estimated in
the range of 150 - 270 MeV (this corresponds to the broad range
of the critical energy density $\epsilon_{c} \simeq 0.26-5.5$
GeV/fm$^{-3}$). The upper limit belongs to a pure $SU(3)$ theory,
whereas adding quarks causes lowering of $T_{c}$ even to about
150 MeV ($\epsilon_{c} \simeq 0.26$ GeV/fm$^{-3}$ respectively).
Since NA50 Collaboration estimates for the energy density obtained
in the central rapidity region (CRR) give the value of 3.5
GeV/fm$^{-3}$ for the most central Pb-Pb data point
\cite{Abreu:2000ni}, it is argued that the region of the existence
of the QGP has been reached in Pb-Pb collisions at CERN SPS.

Soon after the announcement, the paper entitled "Evidence for
deconfinement of quarks and gluons from the $J/\Psi$  suppression
pattern measured in Pb-Pb collisions at the CERN-SPS" appeared
\cite{Abreu:2000ni}. As the title suggests, the main argument for
the QGP creation during Pb-Pb collisions at the CERN SPS is the
observation of the suppression of $J/\Psi$ relative yield.
$J/\Psi$ suppression as a signal for the QGP formation was
originally proposed by Matsui and Satz \cite{Matsui:1986dk}. The
clue point of the NA50 Collaboration paper is the figure (denoted
as Fig.6 there), where experimental data for Pb-Pb collision
values of ${B_{\mu\mu}\sigma_{J/\psi}} \over {\sigma_{DY}}$ (the
ratio of the $J/\Psi$ to the Drell-Yan production cross-section
times the branching ratio of the $J/\Psi$ into a muon pair) are
presented together with some conventional predictions. Here,
"conventional" means that $J/\Psi$ suppression is due to $J/\Psi$
absorption in ordinary hadron matter. Since all those
conventional curves saturate at high transverse energy $E_{T}$,
but the experimental data fall from $E_{T} \simeq$ 90 GeV much
lower and this behaviour could be reproduced on the base of
$J/\Psi$ disintegration in QGP \cite{Satz:1999si}, it is argued
that QGP had to appear during most central Pb-Pb collisions. In
general, the immediate reservation about such reasoning is that
besides those already known (see e.g.
\cite{Ftacnik:1988qv,Vogt:1988fj,Gavin:1988hs,Blaizot:1989ec,Vogt:1999cu}
and references [12-15] in \cite{Abreu:2000ni}), there could be a
huge number of different conventional models, so until this
subject is cleared up completely, no one is legitimate to ruled
out the absorption picture of $J/\Psi$ suppression.

In the following talk, the more systematic and general description
of $J/\Psi$ absorption in the framework of statistical analysis
will be presented. The main features of the model are
\cite{Prorok:2000kv}:
\begin{itemize}
\item{1.} a multi-component non-interacting hadron gas appears in
the CRR instead of the QGP;
\item{2.} the gas expands longitudinally and transversely;
\item{3.} $J/\Psi$ suppression is the result of inelastic
scattering on constituents of the gas and on nucleons of
colliding ions.
\end{itemize}

\section {The timetable of events in the CRR}
\label{timetable}

For a given A-B collision $t=0$ is fixed at the moment of the
maximal overlap of the nuclei (for more details see e.g.
\cite{Blaizot:1989ec}). As nuclei pass each other charmonium
states are produced as the result of gluon fusion. After half of
the time the nuclei need to cross each other ($t \sim$ 0.5 fm),
matter appears in the CRR. It is assumed that the matter
thermalizes almost immediately and the moment of thermalization,
$t_{0}$, is estimated at about 1 fm
\cite{Blaizot:1989ec,Bjorken:1983qr}. Then the matter begins its
expansion and cooling and after reaching the freeze-out
temperature, $T_{f.o.}$, it ceases as a thermodynamical system.
The moment when the temperature has decreased to $T_{f.o.}$ is
denoted as $t_{f.o.}$. Since the matter under consideration is
the hadron gas, any phase transition does not take place during
cooling here.

For the description of the evolution of the matter, relativistic
hydrodynamic is explored. The longitudinal component of the
solution of hydrodynamic equations (the exact analytic solution
for an (1+1)-dimensional case) reads (for details see e.g.
\cite{Bjorken:1983qr,Cleymans:1986wb})

\begin{equation}
s(\tau)= { {s_{0}\tau_{0}} \over \tau } \;,\;\;\;\;\;\;\;\;\;
n_{B}(\tau)= { {n_{B}^{0}\tau_{0}} \over \tau }
\;,\;\;\;\;\;\;\;\;\; v_{z}={z \over t} \label{1}
\end{equation}

\noindent where $\tau=\sqrt{t^{2}-z^{2}}$ is a local proper time,
$v_{z}$ is the $z$-component of fluid velocity ($z$ is a collision
axis) and $s_{0}$ and $n_{B}^{0}$ are initial densities of the
entropy and the baryon number respectively. For $n_{B}=0$ and the
uniform initial temperature distribution with the sharp edge at
the border established by nuclei radii, the full solution of
(3+1)-dimensional hydrodynamic equations is known
\cite{Baym:1983sr}. The evolution derived is the decomposition of
the longitudinal expansion inside a slice bordered by the front of
the rarefaction wave and the transverse expansion which is
superimposed outside of the wave. Because small but nevertheless
non-zero baryon number densities are considered here, the
above-mentioned description of the evolution has to be treated as
an assumption in the presented model. The rarefaction wave moves
radially inward with a sound velocity $c_{s}$. The sound velocity
squared is given by $c_{s}^{2}= { {\partial P} \over {\partial
\epsilon} }$ and can be evaluated numerically
\cite{Prorok:1995xz,Prorok:2001ut}.

\section { The multi-component hadron gas }
\label{hadgas}

For an ideal hadron gas in thermal and chemical equilibrium, which
consists of $l$ species of particles, energy density $\epsilon$,
baryon number density $n_{B}$, strangeness density $n_{S}$ and
entropy density $s$ read ($\hbar=c=1$ always)

\begin{mathletters}
\label{eqstate}
\begin{equation}
\epsilon = { 1 \over {2\pi^{2}}} \sum_{i=1}^{l} (2s_{i}+1)
\int_{0}^{\infty}dp\,{ { p^{2}E_{i} } \over { \exp \left\{ {{
E_{i} - \mu_{i} } \over T} \right\} + g_{i} } } \ , \label{energy}
\end{equation}

\begin{equation}
n_{B}={ 1 \over {2\pi^{2}}} \sum_{i=1}^{l} (2s_{i}+1)
\int_{0}^{\infty}dp\,{ { p^{2}B_{i} } \over { \exp \left\{ {{
E_{i} - \mu_{i} } \over T} \right\} + g_{i} } } \ ,
\label{barnumb}
\end{equation}

\begin{equation}
n_{S}={1 \over {2\pi^{2}}} \sum_{i=1}^{l} (2s_{i}+1)
\int_{0}^{\infty}dp\,{ { p^{2}S_{i} } \over { \exp \left\{ {{
E_{i} - \mu_{i} } \over T} \right\} + g_{i} } } \ ,
\label{strange}
\end{equation}

\begin{equation}
s={1 \over {6\pi^{2}T^{2}} } \sum_{i=1}^{l} (2s_{i}+1)
\int_{0}^{\infty}dp\, { {p^{4}} \over { E_{i} } } { { (E_{i} -
\mu_{i}) \exp \left\{ {{ E_{i} - \mu_{i} } \over T} \right\} }
\over { \left( \exp \left\{ {{ E_{i} - \mu_{i} } \over T} \right\}
+ g_{i} \right)^{2} } }\ , \label{entropy}
\end{equation}
\end{mathletters}

\noindent where $E_{i}= ( m_{i}^{2} + p^{2} )^{1/2}$ and $m_{i}$,
$B_{i}$, $S_{i}$, $\mu_{i}$, $s_{i}$ and $g_{i}$ are the mass,
baryon number, strangeness, chemical potential, spin and a
statistical factor of specie $i$ respectively (an antiparticle is
treated as a different specie). And $\mu_{i} = B_{i}\mu_{B} +
S_{i}\mu_{S}$, where $\mu_{B}$ and $\mu_{S}$ are overall baryon
number and strangeness chemical potentials respectively.

To obtain the time dependence of temperature and baryon number and
strangeness chemical potentials one has to solve numerically
equations (\ref{barnumb} - \ref{entropy}) with $s$, $n_{B}$ and
$n_{S}$ given as time dependent quantities. For $s(\tau)$ and
$n_{B}(\tau)$ expressions (\ref{1}) are taken and $n_{S}=0$ since
the overall strangeness equals zero during all the evolution (for
more details see \cite{Prorok:1995xz}).

\section {$J/\Psi$ absorption in the expanding hadronic gas }
\label{absorb}

As it has been already mentioned in Sect.~\ref{timetable}
charmonium states are produced in the beginning of the collision,
when nuclei overlap. For simplicity, it is assumed that the
production of $c\bar{c}$ states takes place at $t=0$. To describe
$J/\Psi$ absorption quantitatively, the idea of
\cite{Blaizot:1989ec} is generalized to the case of the
multi-component massive gas, here. Since in the CRR longitudinal
momenta of particles are much lower than transverse ones (in the
c.m.s. frame of nuclei), $J/\Psi$ longitudinal momentum is put at
zero. Additionally, only the plane $z=0$ is under consideration,
here. For the simplicity of the model, it is assumed that all
charmonium states are completely formed and can be absorbed by the
constituents of a surrounding medium from the moment of creation.
The absorption is the result of a $c\bar{c}$ state inelastic
scattering on the constituents of the hadron gas through
interactions of the type

\begin{equation}
c\bar c+h \longrightarrow D+\bar D+X\\ ,  \label{psiabs}
\end{equation}


\noindent where $h$ denotes a hadron, $D$ is a charm meson and $X$
means a particle which is necessary to conserve the charge, baryon
number or strangeness.

\begin{figure}
\begin{center}{
{\epsfig{file=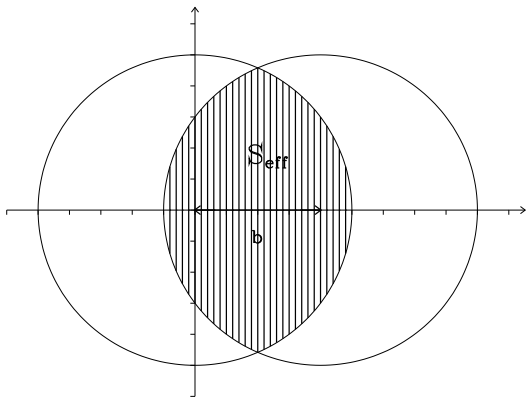,width=7cm}} }\end{center} \caption{View
of a Pb-Pb collision at impact parameter $b$ in the transverse
plane ($z=0$). The region where the nuclei overlap is hatched and
denoted by $S_{eff}$.} \label{Fig.1.}
\end{figure}

Since the most crucial for the problem of QGP existence are Pb-Pb
collisions (see remarks in Sect.~\ref{intro}), the further
considerations are done for this case. So, for the Pb-Pb
collision at impact parameter $b$, the situation in the plane
$z=0$ is presented in Fig.\,\ref{Fig.1.}, where $S_{eff}$ means
the area of the overlap of the colliding nuclei.

Additionally, it is assumed that the hadron gas, which appears in
the space between the nuclei after they have crossed each other,
also has the shape of $S_{eff}$ at $t_{0}$ in the plane $z=0$.
Then, the transverse expansion starts in the form of the
rarefaction wave moving inward $S_{eff}$ at $t_{0}$.

From the considerations based on the relativistic kinetic
equation (for details see \cite{Prorok:2000kv,Blaizot:1989ec}),
the survival fraction of $J/\Psi$ in the hadron gas as a function
of the initial energy density $\epsilon_{0}$ in the CRR is
obtained:

\begin{equation}
{\cal N}_{h.g.}(\epsilon_{0}) = \int dp_{T}\;
g(p_{T},\epsilon_{0}) \cdot \exp \left\{ -\int_{t_{0}}^{t_{final}}
dt \sum_{i=1}^{l} \int { {d^{3}\vec{q}} \over {(2\pi)^{3}} }
f_{i}(\vec{q},t) \sigma_{i} v_{rel,i} { {p_{\nu}q_{i}^{\nu}} \over
{EE^{\prime}_{i}} } \right\}\ , \label{surhg}
\end{equation}

\noindent where the sum is over all taken species of scatters
(hadrons), $p^{\nu}=(E,\vec{p}_{T})$ and
$q_{i}^{\nu}=(E^{\prime}_{i},\vec{q})$ are four momenta of
$J/\Psi$ and hadron specie $i$ respectively, $\sigma_{i}$ states
for the absorption cross-section of $J/\Psi-h_{i}$ scattering,
$v_{rel,i}$ is the relative velocity of $h_{i}$ hadron with
respect to $J/\Psi$ and $M$ and $m_{i}$ denote $J/\Psi$ and
$h_{i}$ masses respectively ($M= 3097$ MeV). The function
$g(p_{T},\epsilon_{0})$ is the $J/\Psi$ initial momentum
distribution. It has a gaussian form and reflects gluon multiple
elastic scatterings on nucleons before their fusion into a
$J/\Psi$ in the first stage of the collision
\cite{Hufner:1988wz,Gavin:1988tw,Blaizot:1989hh,Gupta:1992cd}.
The upper limit of the integration over time in (\ref{surhg}),
namely $t_{final}$ is the minimal value of $\langle
t_{esc}\rangle$ and $t_{f.o.}$. The quantity $\langle
t_{esc}\rangle$ is the average time of the escape of $J/\Psi$'s
from the hadron medium for given values of $b$ and $J/\Psi$
velocity $\vec{v}= \vec{p}_{T}/E$. Note that the average is taken
with the weight

\begin{equation}
p_{J/\Psi}(\vec{r}) = { {T_{A}(\vec{r})T_{B}(\vec{r} - \vec{b})}
\over {T_{AB}(b)} } \;, \label{weight}
\end{equation}

\noindent where $T_{A}(\vec{s}) = \int dz \rho_{A}(\vec{s},z)$
and $\rho_{A}(\vec{s},z)$ is the nuclear matter density
distribution. For the last quantity, the Woods-Saxon nuclear
matter density distribution with parameters from \cite{Jager} is
taken. In the integration over hadron momentum in (\ref{surhg})
the threshold for the reaction (\ref{psiabs}) is included, i.e.
$\sigma_{i}$ equals zero for $(p^{\nu}+q_{i}^{\nu})^{2} < (2m_{D}
+ m_{X})^{2}$ and is constant elsewhere ($m_{D}$ is a charm meson
mass, $m_{D}= 1867$ MeV). Also the usual Bose-Einstein or
Fermi-Dirac distribution for hadron specie $i$ is used in
(\ref{surhg})

\begin{equation}
f_{i}(\vec{q},t)=f_{i}(q,t)={ {2s_{i}+1} \over { \exp \left\{ { {
E^{\prime}_{i}-\mu_{i}(t)} \over {T(t)} } \right\} + g_{i} } }\ .
\label{BoseF}
\end{equation}

As far as $\sigma_{i}$ is concerned, there are no data for every
particular $J/\Psi-h_{i}$ scattering. Therefore, only two types of
the cross-section, the first, $\sigma_{b}$, for $J/\Psi-baryon$
scattering and the second, $\sigma_{m}$, for $J/\Psi-meson$
scattering are assumed, here. In the model presented $\sigma_{b}$
is put at $\sigma_{J/\psi N}$ --- ~ $J/\Psi-Nucleon$ absorption
cross-section. And values of $\sigma_{J/\psi N}$ in the range of
3-5 mb are taken from p-A data
\cite{Gerschel:1988wn,Badier:1983dg,Gavin:1997yd}. The
$J/\Psi-meson$ cross-section $\sigma_m$ is assumed to be 2/3 of
$\sigma_{b}$, which is due to the quark counting.

As it has been already suggested \cite{Gerschel:1988wn} also
$J/\Psi$ scattering in nuclear matter should be included in any
$J/\Psi$ absorption model. This could be done with the
introduction of $J/\Psi$ survival factor in nuclear matter
\cite{Gavin:1997yd,Gerschel:1992uh,Gavin:1996en,Bella}

\begin{equation}
{\cal N}_{n.m.}(\epsilon_{0}) \cong \exp \left\{ -\sigma_{J/\psi
N} \rho_{0} L \right\}\ , \label{surnm}
\end{equation}

\noindent where $\rho_{0}$ is the nuclear matter density and $L$
the mean path length of the $J/\Psi$ through the colliding
nuclei. The length $L$ is expressed by the following formula
\cite{Bella}:

\begin{equation}
\rho_{0}L(b) = {1 \over {2 T_{AB}} } \int d^{2}\vec{s}\;
T_{A}(\vec{s}) T_{B}(\vec{s} - \vec{b}) \left[ {{A-1} \over A}
T_{A}(\vec{s}) + {{B-1} \over B} T_{B}(\vec{s} - \vec{b})
\right]\ , \label{length}
\end{equation}

\noindent where $T_{AB}(b) = \int d^{2}\vec{s}\; T_{A}(\vec{s})
T_{B}(\vec{s} - \vec{b})$.

Since $J/\Psi$ absorptions: in nuclear matter and in the hadron
gas, are separate in time, $J/\Psi$ survival factor for a
heavy-ion collision with the initial energy density
$\epsilon_{0}$, could be defined as

\begin{equation}
{\cal N}(\epsilon_{0}) = {\cal N}_{n.m.}(\epsilon_{0}) \cdot {\cal
N}_{h.g.}(\epsilon_{0})\ . \label{surv}
\end{equation}

\noindent Note that since right sides of (\ref{surhg}) and
(\ref{surnm}) include parts which depend on impact parameter $b$
and the left sides are functions of $\epsilon_{0}$ only, the
expression converting the first quantity to the second (or
reverse) should be given. This is done with the use of the
dependence of $\epsilon_{0}$ on the transverse energy $E_{T}$
extracted from NA50 data \cite{Abreu:2000ni} (for details see
\cite{Prorok:2000kv}).

To make the model as much realistic as possible, one should keep
in mind that only about $60 \%$ of $J/\Psi$'s measured are
directly produced during collision. The rest is the result of
$\chi$ ($\sim 30 \%$) and $\psi'$ ($\sim 10 \%$) decay
\cite{Satz:1997ib}. Therefore the realistic $J/\Psi$ survival
factor could be expressed as

\begin{equation}
{\cal N}(\epsilon_{0})=0.6{\cal
N}_{J/\psi}(\epsilon_{0})+0.3{\cal N}_{\chi}(\epsilon_{0})+
0.1{\cal N}_{\psi'}(\epsilon_{0})\;, \label{survsum}
\end{equation}

\noindent where ${\cal N}_{J/\psi}(\epsilon_{0})$, ${\cal
N}_{\chi}(\epsilon_{0})$ and ${\cal N}_{\psi'}(\epsilon_{0})$ are
given also by formulae (\ref{surhg}), (\ref{surnm}) and
(\ref{surv}) but with $g(p_{T},\epsilon_{0})$, $\sigma_{J/\psi
N}$ and $M$ changed appropriately (for details see
\cite{Prorok:2000kv}).

To complete, also values of cross-sections for $\chi-baryon$ and
$\psi'-baryon$ scatterings are needed. For simplicity, it is
assumed that both these cross-sections are equal to
$J/\Psi-baryon$ one.  Since $J/\Psi$ is smaller than $\chi$ or
$\psi'$, this means that $J/\Psi$ suppression evaluated according
to (\ref{survsum}) is {\it underestimated} here.

\section { Results }

To calculate formula (\ref{surhg}) initial values $s_{0}$ and
$n_{B}^{0}$ are needed. For the estimation of the last quantity
experimental results for S-S \cite{Baechler:1991pd} and Au-Au
\cite{Stachel:1999rc,Ahle:1998jc} collisions are explored. The
S-S data give $n_{B}^{0} \cong 0.25$ fm$^{-3}$ and Au-Au data
$n_{B}^{0} \cong 0.65$ fm$^{-3}$. These values are for central
collisions, for more peripheral ones the initial baryon number
density should be lower. To simplify numerical calculations,
$n_{B}^{0}$ is kept constant over the whole range of $b$. But, to
check the possible dependence on $n_{B}^{0}$, also the much lower
value $n_{B}^{0} = 0.05$ fm$^{-3}$ is examined. To find $s_{0}$,
(\ref{energy} - \ref{strange}) with respect to $T$, $\mu_{S}$ and
$\mu_{B}$ are solved. Then, having put $T_{0}$, $\mu_{S}^{0}$ and
$\mu_{B}^{0}$ into (\ref{entropy}), $s_{0}$ is obtained. The last
parameter of the model is the freeze-out temperature. Two values
$T_{f.o.}=100, 140$ MeV are taken here and they agree well with
estimates based on hadron yields \cite{Stachel:1999rc}.

\begin{figure}[htb]
\begin{minipage}[t]{75mm}
{\psfig{figure= 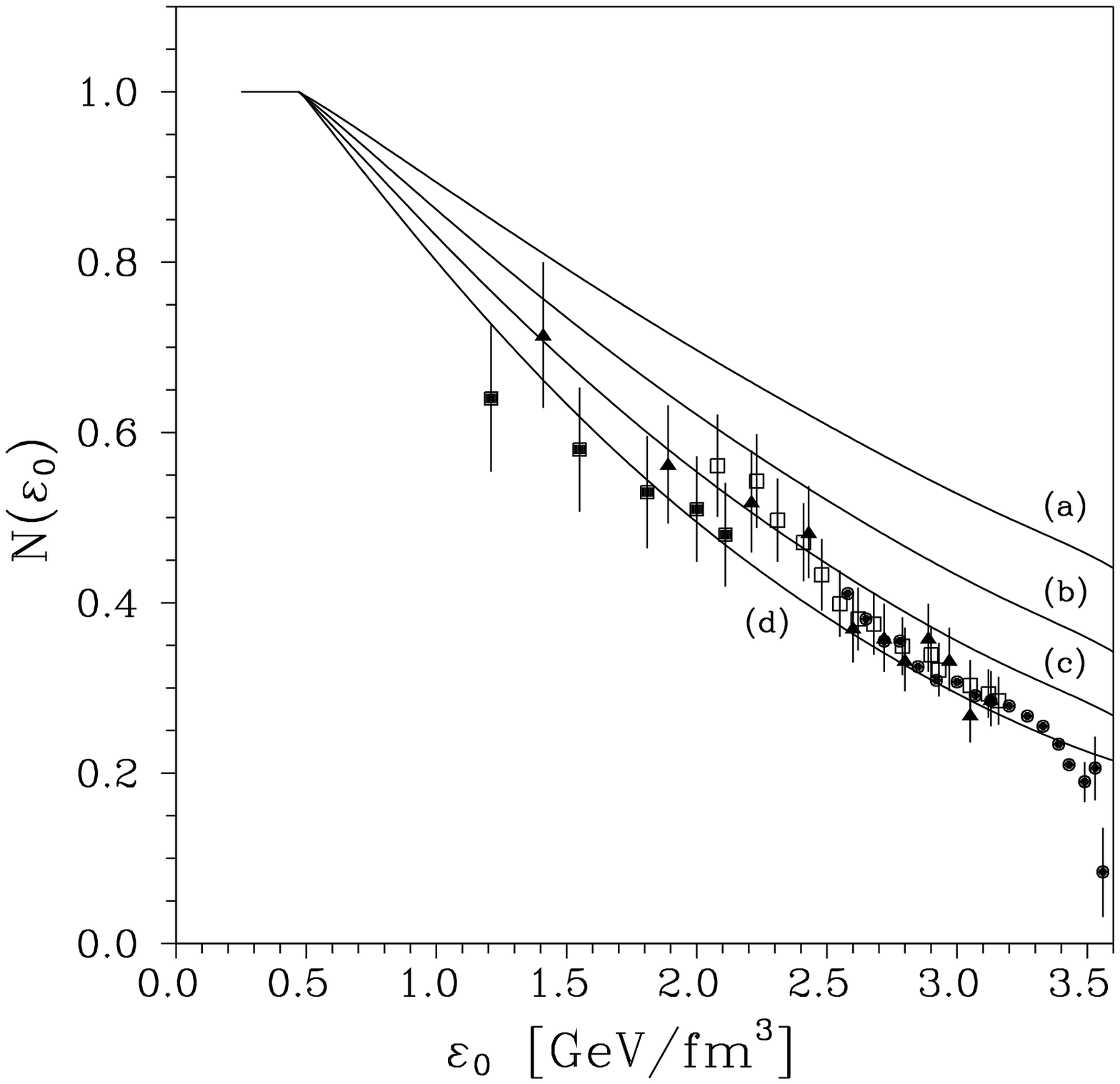,height=\textwidth,width=\textwidth}}
\caption{$J/\Psi$ suppression in the longitudinally expanding
hadron gas for $n_{B}^{0}=0.25$ fm$^{-3}$ and $T_{f.o.}=140$ MeV:
(a) $\sigma_{b}=3$ mb, $\sigma_{m}=2$ mb; (b) $\sigma_{b}= 4$ mb,
$\sigma_{m}=2.66$ mb; (c) $\sigma_{b}=5$ mb, $\sigma_{m}=3.33$ mb;
(d) $\sigma_{b}=6$ mb, $\sigma_{m}=4$ mb . The black squares
correspond to the NA38 S-U data \protect\cite{Abr}, the black
triangles correspond to the 1996 NA50 Pb-Pb data, the white
squares to the 1996 analysis with minimum bias and the black
points to the 1998 analysis with minimum bias
\protect\cite{Abreu:2000ni}. } \label{Fig.2.}
\end{minipage}
\hspace{\fill}
\begin{minipage}[t]{75mm}
{\psfig{figure= 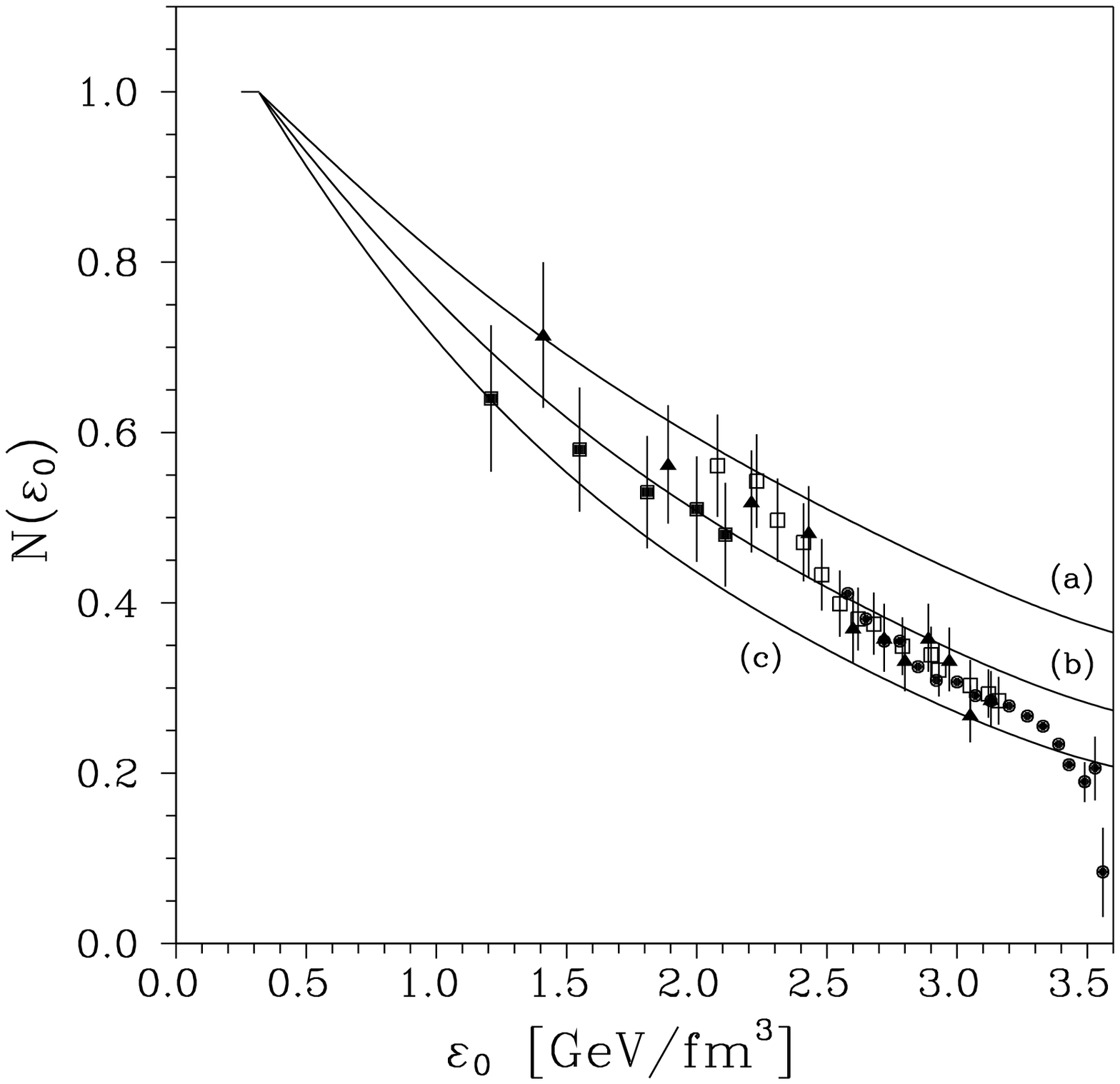 ,height=\textwidth,width=\textwidth}}
\caption{Same as Fig.\,\ref{Fig.2.} (except case (d), which is not
presented here) but for $T_{f.o.}=100$ MeV.} \label{Fig.3.}
\end{minipage}
\end{figure}

Firstly, formula (\ref{survsum}) with the use of (\ref{surhg})
only instead of (\ref{surv}) is calculated for the case of
(1+1)-dimensional expansion and results are presented in
Figs.\,\ref{Fig.2.}-\ref{Fig.3.}.

For comparison, also the experimental data are shown in
Figs.\,\ref{Fig.2.}-\ref{Fig.3.}. The experimental survival
factor is defined as

\begin{equation}
{\cal N}_{exp}={ { {B_{\mu\mu}\sigma_{J/\psi}^{AB}}\over
{\sigma_{DY}^{AB}} } \over { {B_{\mu\mu}\sigma_{J/\psi}^{pp}}
\over {\sigma_{DY}^{pp}}  } }\;\;, \label{survexp1}
\end{equation}

\noindent where ${B_{\mu\mu}\sigma_{J/\psi}^{AB(pp)}} \over
{\sigma_{DY}^{AB(pp)}}$ is the ratio of the $J/\Psi$ to the
Drell-Yan production cross-section in A-B(p-p) interactions times
the branching ratio of the $J/\Psi$ into a muon pair. The values
of the ratio for p-p, S-U and Pb-Pb are taken from
\cite{Abreu:2000ni,Abr,Abr50}. The results of numerical
evaluations of (\ref{survsum}) agree with the data well
qualitatively for greater baryonic cross-section $\sigma_{b}$ and
(or) for the lower freeze-out temperature $T_{f.o.}$, except the
last point measured for the highest $E_{T}$ bin.

\begin{figure}[htb]
\begin{minipage}[t]{75mm}
{\psfig{figure= 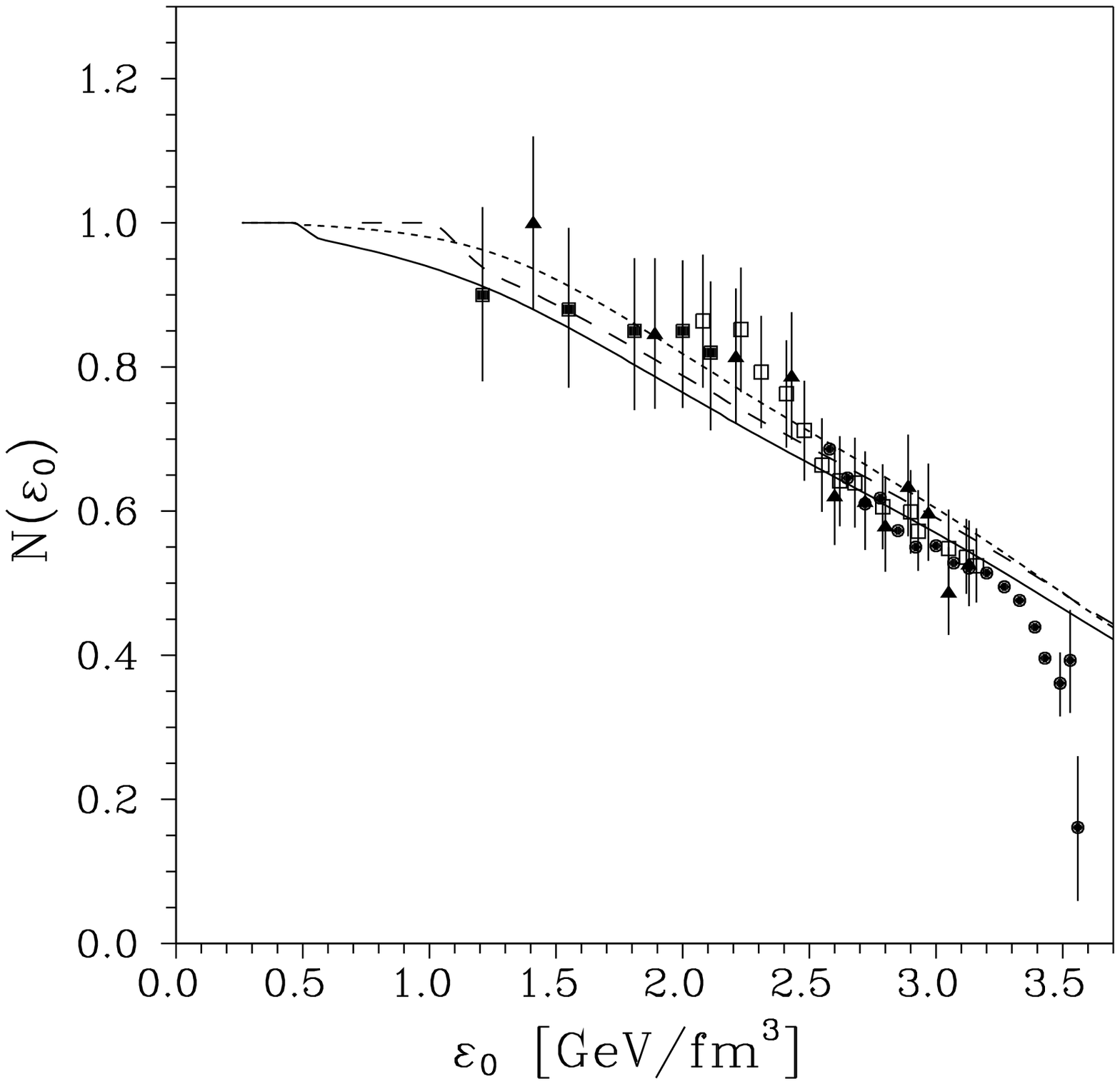,height=\textwidth,width=\textwidth}}
 \caption{$J/\Psi$ suppression in the longitudinally and
transversely expanding hadron gas for the Woods-Saxon nuclear
matter density distribution and $\sigma_{b}=4$ mb,
$\sigma_{m}=2.66$ mb and $T_{f.o.}=140$ MeV. The curves correspond
to $n_{B}^{0}=0.25$ fm$^{-3}$, $c_{s}=0.45$, $r_{0}=1.2$ fm
(solid), $n_{B}^{0}=0.65$ fm$^{-3}$, $c_{s}=0.46$, $r_{0}=1.2$ fm
(dashed) and  $n_{B}^{0}=0.25$ fm$^{-3}$, $c_{s}=0.45$,
$r_{0}=1.12$ fm (short-dashed). The black squares represent the
NA38 S-U data \protect\cite{Abr}, the black triangles represent
the 1996 NA50 Pb-Pb data, the white squares the 1996 analysis with
minimum bias and the black points the 1998 analysis with minimum
bias \protect\cite{Abreu:2000ni}, but the data are "cleaned out"
from the contribution of $J/\Psi$ scattering in the nuclear
matter in accordance with (\ref{survgas}).} \label{Fig.4.}
\end{minipage}
\hspace{\fill}
\begin{minipage}[t]{75mm}
{\psfig{figure= 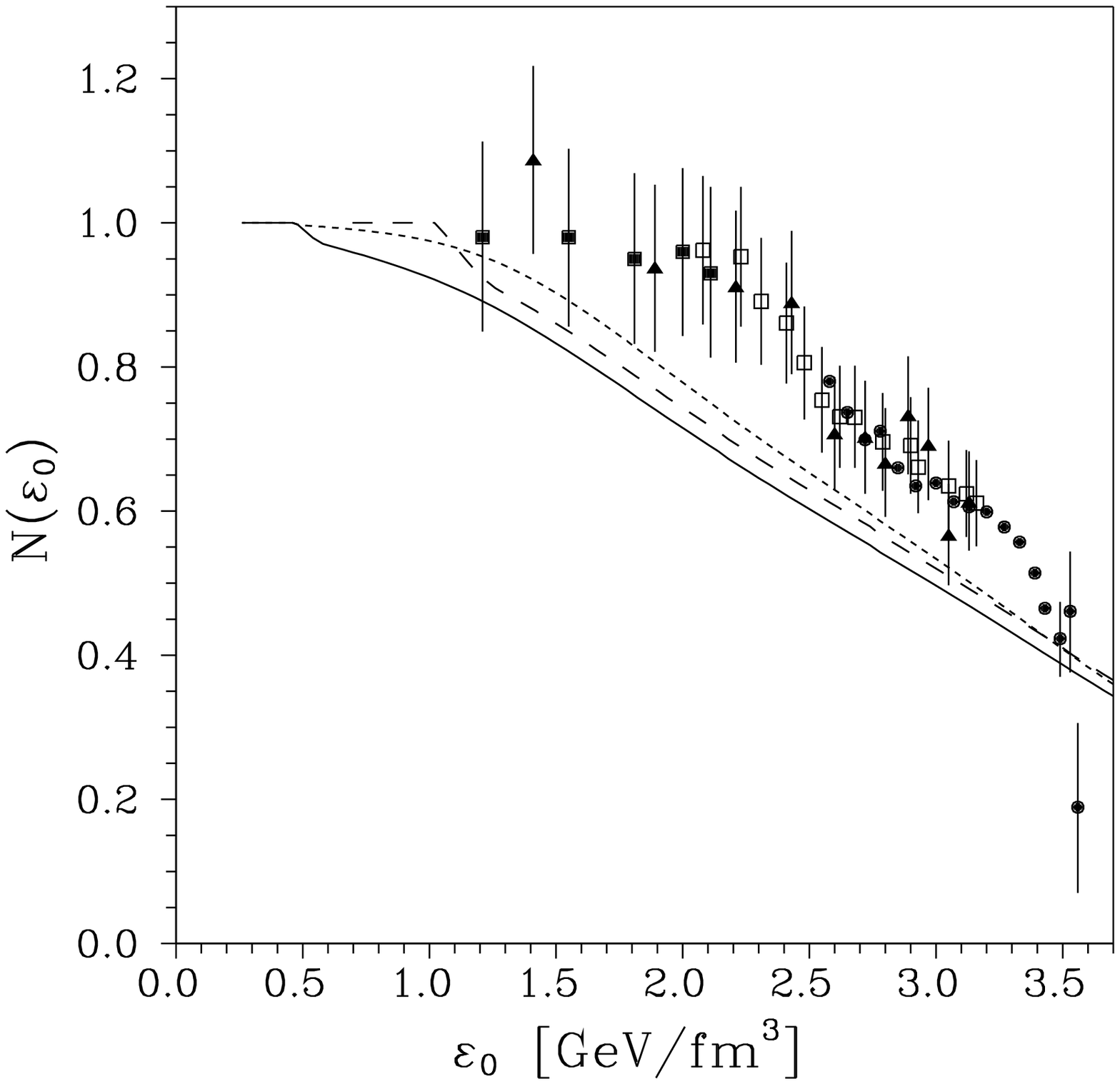 ,height=\textwidth,width=\textwidth}}
 \caption{Same as Fig.\,\ref{Fig.4.} but for $\sigma_{b}=5$ mb and
$\sigma_{m}=3.33$ mb. } \label{Fig.5.}
\end{minipage}
\end{figure}

Now the full (3+1)-dimensional hydrodynamic evolution (in the form
described in Sect.~\ref{timetable}) and $J/\Psi$ absorption in
nuclear matter will be taken into account and results are
depicted in Figs.\,\ref{Fig.4.}-\ref{Fig.5.}. To draw also S-U
data together with Pb-Pb ones, instead of multiplying ${\cal
N}_{h.g.}$ by ${\cal N}_{n.m.}$, ~ ${\cal N}_{exp}$ is divided by
${\cal N}_{n.m.}$, i.e. "the experimental $J/\Psi$ hadron gas
survival factor" is defined as

\begin{equation}
\tilde{{\cal N}}_{exp}= \exp \left\{ \sigma_{J/\psi N} \rho_{0} L
\right\} \cdot {\cal N}_{exp}\;, \label{survgas}
\end{equation}

\noindent and values of this factor are drawn in
Figs.\,\ref{Fig.4.}-\ref{Fig.5.} as the experimental data. The
quantity $r_{0}$ is a parameter from the expression for a nucleus
radius $R_{A} = r_{0} \cdot A^{{1 \over 3}}$. The value of
$R_{A}$ is used to fix the initial position of the rarefaction
wave in the evaluation of $\langle t_{esc}\rangle$ in
(\ref{surhg}). It has turned out that the case of
$n_{B}^{0}=0.05$ fm$^{-3}$ does not differ substantially from
that of $n_{B}^{0}=0.25$ fm$^{-3}$, so curves for
$n_{B}^{0}=0.05$ fm$^{-3}$ are not depicted in
Figs.\,\ref{Fig.4.}-\ref{Fig.5.}.

Having compared Figs.\,\ref{Fig.4.}-\ref{Fig.5.} with
Figs.\,\ref{Fig.2.}-\ref{Fig.3.}, one can see that adding the
transverse expansion changes the final (theoretical) pattern of
$J/\Psi$ suppression qualitatively. Now the curves for the case
including the transverse expansion are not convex, in opposite to
the case with the longitudinal expansion only, where the curves
are. As far as the pattern of suppression is concerned,
theoretical curves do not fall steep enough at high
$\epsilon_{0}$ to cover the data area. But for some choice of
parameters, namely for $\sigma_{b}$ somewhere between 4 and 5 mb
and for $r_{0}=1.12$ fm, a quite satisfactory curve could have
been obtained. Precisely, again only the highest $E_{T}$ bin
point falls down of the range of theoretical estimates completely.
But the error bar of this point is very wide. And also the
contradiction in positions of the last three points of the 1998
data can be seen. This means that the high $E_{T}$ region should
be measured once more with the better accuracy to state
definitely whether the abrupt fall of the experimental survival
factor takes place or not there.

\begin{figure}
\begin{center}{
{\epsfig{file=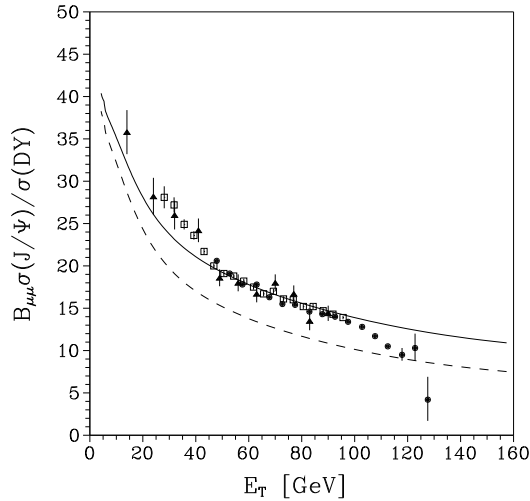,width=7cm}} }\end{center}
\caption{$J/\Psi$ survival factor times
${B_{\mu\mu}\sigma_{J/\psi}^{pp}} \over {\sigma_{DY}^{pp}}$ in
the longitudinally and transversely expanding hadron gas for the
Woods-Saxon nuclear matter density distribution and
$n_{B}^{0}=0.25$ fm$^{-3}$, $T_{f.o.}=140$ MeV, $c_{s}=0.45$ and
$r_{0}=1.2$ fm. The curves correspond to $\sigma_{b}=4$ mb
(solid) and $\sigma_{b}=5$ mb (dashed). The black triangles
represent the 1996 NA50 Pb-Pb data, the white squares the 1996
analysis with minimum bias and the black points the 1998 analysis
with minimum bias \protect\cite{Abreu:2000ni}. } \label{Fig.6.}
\end{figure}

To support the conclusion, main results from
Figs.\,\ref{Fig.4.}-\ref{Fig.5.} are presented in
Fig.\,\ref{Fig.6.} again. The original data \cite{Abreu:2000ni}
for ${B_{\mu\mu}\sigma_{J/\psi}^{PbPb}} \over
{\sigma_{DY}^{PbPb}}$ and $J/\Psi$ survival factors given by
(\ref{survsum}) multiplied by ${B_{\mu\mu}\sigma_{J/\psi}^{pp}}
\over {\sigma_{DY}^{pp}}$ and now as functions of $E_{T}$ are
depicted there (for details see \cite{Prorok:2000kv}).

\noindent As it has been already mentioned, the main disagreement
with the data appears in the last experimental point of the 1998
analysis.

The $charmonium-baryon$ inelastic cross-section is the most
crucial parameter in the model presented. So to be sure what is
the exact result of $J/\Psi$ absorption, one should know how this
cross-section behaves in the hot hadron environment. The newest
estimations of $\pi+J/\Psi$, $\rho+J/\Psi$ and $J/\Psi+N$
cross-sections at high invariant collision energies
\cite{Tsushima:2000cp,Sibirtsev:2000aw} give values of
$\sigma_{b}$ and $\sigma_{m}$ of the same order as assumed here.
Also it should be stressed that the $charmonium-hadron$ inelastic
cross-sections are considered as constant quantities in the model
presented. But, as the results of just mentioned papers suggest,
they are not constant at all. The cross-sections are growing
functions of the invariant collision energy $\sqrt{s}$.
Therefore, one could think naively that the increase of
$\epsilon_{0}$ (or in other words $E_{T}$) causes the increase of
the invariant collision energy $\sqrt{s}$ on the average and
further the increase of the $charmonium-hadron$ inelastic
cross-sections. In this way, the line describing $J/\Psi$ survival
factor could be placed close to the solid curve of
Fig.\,\ref{Fig.6.} for low $\epsilon_{0}$ ($E_{T}$), and then, as
the $charmonium-hadron$ inelastic cross-sections increase, this
line would go closer to the dashed curve of Fig.\,\ref{Fig.6.}
for high $\epsilon_{0}$ ($E_{T}$). So, the experimental pattern
of $J/\Psi$ suppression could be recovered.

As the last remark, the author would like to call reader's
attention to Fig.5 of \cite{Wong:2001if}. The solid and dotted
curves in that figure are exactly the same as curves just
presented in Fig.\,\ref{Fig.6.}. But the distortion in the QGP is
the main reason for $J/\Psi$ suppression in \cite{Wong:2001if}!
It is also stated there, that results shown "provide evidence for
the production of the quark-gluon plasma in central high-energy
Pb-Pb collisions".

Therefore, the final conclusion is the following: the NA50 PB-PB
data provides evidence for the production of the thermalized and
{\it confined} hadron gas in the central rapidity region of a
PB-PB collision. This implies that $J/\Psi$ suppression is not a
good signal for the quark-gluon plasma appearance at a central
heavy-ion collision.

\end{document}